\documentclass[a4paper]{article}
\usepackage{latexsym}
\begin{document}

\newtheorem{definition}{Definition}
\newtheorem{theorem}{Theorem}
\newtheorem{proposition}{Proposition}
\newtheorem{remark}{Remark}
\newtheorem{corollary}{Corollary}
\newtheorem{lemma}{Lemma}
\newtheorem{observation}{Observation}

\newcommand{\qed}{\hfill$\Box$\medskip}

\title{A Note on Nested String Replacements} 
\author{Holger Petersen\\ 
Reinsburgstr. 75\\
70197 Stuttgart\\
Germany} 

\maketitle

\begin{abstract}
We investigate the number of nested string replacements required to reduce
a string of identical characters to one character. 
\end{abstract}

\section{Introduction}
As part of a test data management project, every 
sequence of digits representing a number in certain strings stored
in a data base had to be replaced by the 
single digit ``1''. This can easily be accomplished by replacing substrings matching 
$[0-9]+$ (representing any non-empty sequence of digits as an 
extended regular expression \cite{ere}) with $1$. 
It however turned out that regular expression matching is 
very slow and that the syntax of functions making use of regular expressions varies between 
data base systems. 
Therefore it is good practice to manipulate strings using efficient and portable 
SQL-functions whenever possible. 

The strings in question had a length of at most 32 characters. Therefore applying five nested 
{\tt REPLACE}-functions each replacing $11$ with $1$ to an initial expression
$$\mbox{\tt TRANSLATE}(s, \mbox{\mbox{'}}023456789\mbox{\mbox{'}}, \mbox{'}111111111\mbox{'})$$
would transform any sequence of digits in string $s$ into $1$.%
\footnote{The function {\tt TRANSLATE} substitutes single characters of its first argument, 
mapping each character appearing in its second argument to the corresponding character of the 
third argument. Characters in the 
second argument whithout a corresponding character in the third argument are removed 
(we will not use this feature). All remaining characters are not modified. The function {\tt REPLACE}
substitutes its third argument for all occurrences of the second in its first argument. Notice that
{\tt REPLACE} searches in a left-to-right manner and continues its search after a substited string. 
See \cite{translate} for further explanations and examples of {\tt TRANSLATE} and {\tt REPLACE}.}

After some experiments, a solution using four nested {\tt REPLACE}-functions was found by 
in turn replacing $1111$, $111$, $11$, and again $11$ with $1$.
Up to 34 digits can be reduced to a single $1$ in this way and it is a natural question, 
whether a further improvement is possible.

\section{Results}
\begin{definition}\label{task}
Task $R(m)$ is to replace all non-empty substrings $1^k$ for $1 \le k \le m$ in a string with $1$
using nested {\tt REPLACE}-functions with strings consisting of character $1$.
\end{definition}

\begin{lemma}\label{first}
The inner-most {\tt REPLACE} in a solution of $R(m)$ with minimum nesting for an $m \ge 2$ 
replaces $1^{\ell}$ with $1^r$ for some $\ell\le m$ and $r \ge 1$. 
\end{lemma}
{\bf Proof.} If the replaced string of the inner-most {\tt REPLACE} has at least $m+1$ symbols, the
{\tt REPLACE} will not influence any string of length at most $m$ and can be omitted from 
a solution of $R(m)$. If the substituted string is empty, at least one input of length at most $m$
(namely $1^{\ell}$) is completely erased and $R(m)$ cannot be solved.
 \qed

\begin{lemma}\label{last}
The outer-most {\tt REPLACE} in a solution of $R(m)$ with minimum nesting for $m \ge 2$ replaces $1^{\ell}$
with $1^r$ for some $\ell\ge 2$ and $r \le 1$.
\end{lemma}
{\bf Proof.} If the replaced string of the outer-most {\tt REPLACE}
consists of one symbol, then the substituted string cannot be empty or have a length greater than one. 
Therefore such a {\tt REPLACE} leaves the text unchanged and would be redundant. 

If the substituted string consists of more than one symbol, the {\tt REPLACE} 
cannot be applied (since would not map to $1$) and again woulkd be redundant.  \qed

\begin{proposition}\label{lower_one}
Task $R(3)$  cannot be solved with one {\tt REPLACE}.
\end{proposition}
{\bf Proof.} By Lemmas~\ref{first} and \ref{last} applied to the single {\tt REPLACE} we only have 
to consider replacing $11$ or $111$ with $1$. Either $111$ or $11$ would be mapped to $11$, 
which shows the claim. \qed

\begin{proposition}\label{lower_two}
Task $R(5)$ cannot be solved with two nested {\tt REPLACE}-functions.
\end{proposition}
{\bf Proof.} Let the nested functions of a hypothetical solution of $R(5)$ be
$$\mbox{\tt REPLACE}(\mbox{\tt REPLACE}(s, \mbox{'}1^{\ell_1}\mbox{'}, \mbox{'}1^{r_1}\mbox{'}), 
\mbox{'}1^{\ell_2}\mbox{'}, \mbox{'}1^{r_2}\mbox{'}),$$
where $s$ is the input. We will derive a contradiction for each possible choice of parameters
$\ell_1$, $r_1$, $\ell_2$, and $r_2$.

By Lemma~\ref{first} we have $\ell_1\le 5$ and $r_1\ge 1$ and by Lemma~\ref{last} we have
$\ell_2\ge 2$ and $r_2\le 1$.

If $\ell_1\ge 4$, the strings $1$, $11$, and $111$ are unchanged by the inner {\tt REPLACE}. Then the
outer {\tt REPLACE} would have to map these strings to $1$, which would be a solution of 
$R(3)$ with one {\tt REPLACE} contradicting 
Proposition~\ref{lower_one}. We therefore only have to consider $1 \le \ell_1 \le 3$.

Let us assume $\ell_1 = 1$. If in addition $r_1 = 1$, the {\tt REPLACE} would be redundant. 
Therefore $r_1 \ge 2$. The outer
{\tt REPLACE} maps $1^{r_1}$ directly to $1$ in order to
handle the input $1$ or erases $1^d$ for a divisor $d \ge 2$ of ${r_1}-1$ leaving a remainder
of one.
In the former case $1^{2r_1}$ as the result of the inner {\tt REPLACE} on
input $11$ would be mapped to $11$. In the latter case
$1^{2r_1}$ would be erased if $d = 2$ or mapped to $11$ if $d \ge 3$.
In each of these cases $R(5)$ is not solved.

If $\ell_1 = 2$, strings $1111$ and $11111$ are mapped to $1^{2r_1}$ and $1^{2r_1+1}$. Both of these
strings are then mapped to $1$ by the outer {\tt REPLACE}. 
If $r_2 = 0$ then $2r_1$ is divisible by $\ell_2\ge 2$ in order to leave a single $1$ from $1^{2r_1+1}$.
But then $1^{2r_1}$ is mapped to the empty string. If $r_2 = 1$ then $\ell_2 = 2r_1+1 \ge 3$,
since otherwise a string with more than two symbols is generated from $1^{2r_1+1}$. 
But now $1^{2r_1}$ with at least two symbols 
is not modified. In either case we derive a contradiction.

If finallly $\ell_1 = 3$, the input $11$ is not changed by the inner {\tt REPLACE}
and $\ell_2 = 2, r_2 = 1$ in order to avoid the
output $11$. The input $11111$ is mapped to $1^{r_1}11$ with at
least three symbols by the inner {\tt REPLACE} and to a string with
at least two symbols by the outer {\tt REPLACE} again contradicting the assumption.\qed

Notice that the bounds of Propositions~\ref{lower_one} and \ref{lower_two} cannot be improved, since
$$
\mbox{\tt REPLACE}(s, \mbox{'}11\mbox{'}, \mbox{'}1\mbox{'})
$$
and
$$
\mbox{\tt REPLACE}(\mbox{\tt REPLACE}(s, \mbox{'}11\mbox{'}, \mbox{'}1\mbox{'}), \mbox{'}11\mbox{'}, \mbox{'}1\mbox{'})
$$
solve $R(2)$ and $R(4)$ respectively.

\begin{theorem}\label{three}
With three nested {\tt REPLACE}-functions $R(m)$ can be solved for any $m \ge 1$.
\end{theorem}
{\bf Proof.} Since $R(4)$ can be solved with two nested {\tt REPLACE}-functions (and these could
be extended by a redundant {\tt REPLACE}), we only have to consider $m \ge 5$.
The follwing sequence of replacements solves $R(m)$ for $m \ge 5$:
$$\mbox{\tt REPLACE}(\mbox{\tt REPLACE}(\mbox{\tt REPLACE}(s, \mbox{'}1\mbox{'}, \mbox{'}1^{m-1}\mbox{'}), \mbox{'}1^{m}\mbox{'}, \mbox{'}1\mbox{'}), \mbox{'}1^{m-2}\mbox{'}, \mbox{'}\mbox{'}).$$
The inner {\tt REPLACE} blows up a block of $k \ge 1$ ones to length $k(m-1) = (k-1)m + (m-k)$. 
By replacing $m$ symbols with $1$ this is reduced to $(k-1) + (m-k) = m-1$ for $k \le m$. Finally the outer 
{\tt REPLACE} erases all but one symbol. \qed

\section{Discussion}
The somewhat surprising solution in the proof of Theorem~\ref{three} makes essential use of 
increasing the length of the input. If we allow length-decreasing replacements only, each 
{\tt REPLACE} maps its input to strings covering a consecutive range of lengths and we can assume that the 
string being substituted is $1$. By starting from the optimal 
solution of $R(4)$ with two nested {\tt REPLACE}-functions (even if replacements are not
not necessarily length-decreasing), an induction shows that $R(10)$ and $R(40)$ are the tasks
that can be solved with three and four length-decreasing {\tt REPLACE}-functions respectively. 
The sequence $2, 4, 10, 40$ appears as A159860 in the collection \cite{OEIS}, where the recursive formula
 $$a(n) = a(n - 1)(a(n - 1) + 6)/4$$
due to N.~Sato is given for the maximum length $a(n)$ of a string of identical characters 
reducible to length one with
$n$ nested replacements (apparently length-decreasing in view of Theorem~\ref{three}). 
The solution described in the Introduction is thus optimal with respect to 
nested {\tt REPLACE}-functions under 
the additional assumption that all replacements are length-decreasing. 

From a practical point of view the length-decreasing solution is apprroximately 40\% faster
than the one from Theorem~\ref{three}, but the latter is still about twice as fast as
a solution based on a regular expressions.

\end{document}